\documentclass[12pt]{article}
\pdfoutput=1
\usepackage{amssymb}
\usepackage{graphicx}
\usepackage{graphics}
\usepackage{amsmath}

\def\d{\partial}
\def\l{\left(} 
\def\r{\right)}
\def\la{\langle }
\def\ra{\rangle }
\newcommand{\be}{\begin{equation}}
\newcommand{\ee}{\end{equation}}
\newcommand{\ba}{\begin{align}}
\newcommand{\ea}{\end{align}}
\newcommand{\bg}{\begin{gather}}
\newcommand{\eg}{\end{gather}}
\newcommand{\bseq}{\begin{subequations}}
\newcommand{\eseq}{\end{subequations}}

\begin{document}
\title{No Cosmic Rays from Curvature Oscillations during Structure
   Formation with $F(R)$-gravity
}
\author{Dmitry Gorbunov$^{1,2}$, Anna Tokareva$^{1,3}$\\
\mbox{}$^{1}${\small \it Institute for Nuclear Research of Russian Academy of
   Sciences, 117312 Moscow,
   Russia}\\ 
\mbox{}$^{2}${\small \it Moscow Institute of Physics and Technology, 
141700 Dolgoprudny, Russia}\\ 
\mbox{}$^{3}${\small \it Faculty of Physics of Moscow State
   University, 119991 Moscow, Russia}
}
\date{}
\maketitle
\begin{abstract}

The Starobinsky model of modified gravity suggested to explain dark
energy may be also considered in the astrophysical context. Recently
it has been pointed out that in contracting regions curvature
oscillations around the GR value may lead to the production of high
energy particles which contribute to the cosmic ray flux. We revisit
these calculations in the Einstein frame and show that the continuous
approximation for the matter density used in the original calculations
is not valid. We show that this problem is generic in $F(R)$-gravity
models introduced to describe the dark energy. We go beyond the
approximation and find the rate of particle production to be
negligible.

\end{abstract}
\section{Introduction}

Theories of modified gravity have been suggested 
as models capable of
explaining the accelerated expansion of the present Universe. 
The model of $F(R)$-gravity proposed by
Starobinsky \cite{Starobinsky:2007hu}
successfully describes an evolution of the late-time Universe imitating the cosmological constant. The
corresponding action for gravity reads 
\be
\label{action}
S=-\frac{M_{\rm P}^2}{2}\int d^4x \sqrt{-g}\l R+F(R)\r, 
\ee 
where the reduced Planck mass $M_{\rm P}$ is expressed via the Newtonian
gravitational constant $G_N$ as follows $M_{\rm
   P}=\sqrt{8\,\pi/G_N}=2.4\times 10^{18}$\,GeV, and 
\be 
\label{FR}
F(R)=\lambda R_0\left(\left(1+\frac{R^2}{R_0^2}\right)^{-n}
-1\right)-\frac{R^2}{6\, m^2}\,.
\ee
Here $m$ is the maximal mass of additional scalar degree of freedom, 
{\it scalaron}, that appears in $F(R)$ gravity, and 
parameter $R_0$ 
fixes the scale of energy density associated with the cosmological constant $\rho_\Lambda$ at present, when the 
Hubble parameter equals $H_0=1.45\times 10^{-42}$\,GeV,\footnote{This simple
   relation is valid for $\lambda\gg 1$.} 
\be
\label{R_0}
R_0=-\frac{2}{\lambda}\frac{\rho_{\Lambda}}{M_P^2}\,=-
\frac{6\, \Omega_{\Lambda}H_0^2}{\lambda}\;, 
\ee
with $\Omega_\Lambda\equiv \rho_\Lambda/\rho_c$ and critical energy
density $\rho_c$ determined by the Friedman equation, 
\be
\label{Friedman}
3\,M_{\rm P}^2\,H_0^2=\rho_c\,. 
\ee

The main question is how to distinguish $F(R)$-gravity from the
cosmological constant and other Dark Energy models? One way is to
probe the Dark Energy equation of state: $F(R)$-gravity may predict
Dark Energy with time-dependent and even phantom pressure-to-density relation
(see e.g. \cite{Nojiri:2006ri,Sotiriou:2008rp,Nojiri:2010wj} 
for reviews). Another way is to look for
possible observable effects in astrophysics.

$F(R)$-gravity is equivalent to the usual gravity and an additional 
scalar field with a non-trivial potential and
coupling to matter fields. In papers
\cite{Arbuzova:2012su, Arbuzova:2013ina} it was pointed out that, in
space regions with rising matter density, growing curvature oscillations
decay into high energy particles, thus contributing to the Cosmic Ray
spectrum.  These oscillations are oscillations of classical scalaron
field.  In this paper we address a question: where do this scalaron
oscillations come from?  Usually, a scalar field oscillates around
minimum if for some reason it is shifted or pushed from the vacuum. In
papers \cite{Arbuzova:2012su, Arbuzova:2013ina} the initial amplitude
of the oscillations is arbitrary while actually there is no
arbitrariness because the history of the expanding Universe exactly
defines the initial conditions for the scalaron field. Namely, scalaron being massive in the early Universe, decays to the Standard model particles providing the absence of scalaron excitations at the present time. However,
scalarons may be produced through the quantum
processes\,\cite{Gorbunov:2014fwa}. We
show below that numerical estimates in \cite{Arbuzova:2012su, Arbuzova:2013ina}, performed for the adopted there initial
conditions, correspond to very unrealistic situation of huge number
density (and noticeable energy density) of the present scalaron
configuration.

Actually, as it was noticed in \cite{Arbuzova:2012su}, the
oscillations arise even if the field is adiabatically settled in the
vacuum because the scalaron minimum itself moves with rising
density. However, this oscillation source leads to parametrically
smaller amplitudes than those considered in \cite{Arbuzova:2012su,
  Arbuzova:2013ina} for relatively large densities. For matter
densities close to the present dark energy (and for parameters in a
particular region) the oscillations might contribute a noticeable
amount to the energy density of Universe (this situation, if
realistic, needs a special study). However, this energy density does
not release as high energy particles. We revisit calculations of
\cite{Arbuzova:2012su, Arbuzova:2013ina} and obtain that the used
approximation of homogeneous matter and scalaron field does not work
in the region where significant particle production was predicted. We
go beyond this approximations and estimate the realistic flux of
produced cosmic rays to be negligible.  Although we study only a
particular model of $F(R)$ gravity the considered effects are expected
to be generic for the $F(R)$ models aimed at describing the dark
energy \cite{Sotiriou:2008rp}. Such models provide the scalaron
potential whose form depends on the background matter density. This
leads to the scalaron oscillations entering to highly non-linear
regime in which the homogeneous description can be invalid.

\section{Scalaron density in contracting objects}

Classical oscillations of scalaron field $\phi$ may be described in
terms of scalaron condensate and the number density of scalarons $n_\phi$
is defined through the amplitude of oscillations as
\cite{Gorbunov:2011zzc}
\be 
n_\phi=\omega \la\phi^2\ra\,.
\ee
Here $\phi$ is a canonically normalized scalaron field and $\omega$ is
the energy of each particle in the condensate, i.e. the {\it scalaron
effective mass}. It 
depends on the surrounding matter density $\rho_m$ as \cite{Gorbunov:2014fwa}\footnote{This dependence is valid until
$\omega\ll m$. In the opposite limit $\omega\approx m$.}
\be 
\label{scalaron-mass}
\omega=\frac{H_0}{\sqrt{2n\lambda(2n+1)}} 
\l\frac{\rho_m}{\rho_c}\r^{n+1}\l\frac{\lambda}{2 \Omega_{\Lambda}}\r^{n+1/2}.
\ee
While an astrophysical object (i.e. halo) contracts, $\rho_m$ grows. 
In \cite{Arbuzova:2012su, Arbuzova:2013ina} matter density $\rho_m$ 
changes linearly with time as 
$\rho_m=\rho_{m0}(1+t/t_J)$ for $t<t_J$ with $t_J$ being the Jeans
time of contraction.

The scalaron field is related to the scalar curvature and 
may be defined through the derivative $F'(R)\equiv dF/dR$ as
\be 
\label{scalaron}
\phi=\frac{\sqrt{6}}{2}M_P \log (1+F'(R))\approx \frac{\sqrt{6}}{2}M_P
F'(R)\,. 
\ee
The equations of motion for action \eqref{action} provide with the equation of
motion for scalaron. It is convenient to write this equation in terms
of dimensionless variables which first has been done in
\cite{Arbuzova:2010iu}. 
Later calculations in \cite{Arbuzova:2012su, Arbuzova:2013ina} are performed
in terms of variable $\xi$ connected with scalaron field $\phi$ as
follows, 
\be 
\label{variable}
\xi= -\frac{1}{2\lambda n} \l \frac{\rho_{m0}}{M_P^2 R_0}\r^{2n+1}\, 
F'(R)=-\frac{1}{2\lambda n} \l 
\frac{\lambda \rho_{m0}}{2\Omega_{\Lambda} \rho_c}\r^{2n+1}\, F'(R)\,.
\ee
The equation of motion for $\xi$, that defines the time 
dependence of scalaron field,
   reads \cite{Arbuzova:2012su}
\be 
\label{xi_eom}
\xi''+\frac{\rho_m}{\rho_{m0}}-y=0\,,
\ee
where $y=y(\xi)$ is obtained from the equation \footnote{
Here we use the value of $R_0$ defined by \eqref{R_0}; 
in \cite{Arbuzova:2012su} it is taken approximately as
  $R_0=-1/t_U^2$ with $t_U$ being the Universe age. This results in
different dependence of $g$ on $\lambda$ and $\Omega_\Lambda$, as
compared to \cite{Arbuzova:2012su}, which is not important, however.}
\be 
\label{y_eom}
\frac{1}{y^{2n+1}}-g y=\xi ,~~
g\equiv\frac{H_0^2}{2n\lambda m^2}\l\frac{\rho_{m0}}{\rho_c}\r^{2n+2} 
\l\frac{\lambda}{2\Omega_{\Lambda}}\r^{2n+1}\,.
\ee
Primes in \eqref{xi_eom} correspond to the derivatives with respect to 
dimensionless time $\tau\equiv m\sqrt{g}t$, and numerically $g\ll 1$.

The scalaron field $\phi$ may be expressed in terms of function $\xi$ by
using \eqref{scalaron} and \eqref{variable} 
\be 
\label{field}
\phi=\sqrt{6}\lambda n M_P\l\frac{2\Omega_{\Lambda}\rho_c}
{\lambda \rho_{m0}}\r^{2n+1} \,\xi\,.
\ee
As it is shown in the paper \cite{Arbuzova:2012su} when initial
conditions expected in General Relativity are imposed, $y=1$ (which implies for the scalaron to be in the minimum of the potential), 
the amplitude of $\xi$-oscillations becomes
\be 
\label{old-initial}
\delta \xi=(\kappa - y_0')(2n+1)^{3/2}, ~~\kappa=\sqrt{6\lambda n}
\l\frac{\rho_{m0}}{\rho_c}\r^{-n-1/2}\l\frac{\lambda}{2\Omega_{\Lambda}}\r^{-n-1/2}.
\ee
Here (in the limit of small $g$) $y_0'$ is proportional 
to the derivative of curvature $R$ and is 
considered as a free parameter in \cite{Arbuzova:2012su}. Since the
case of $y_0'=\kappa$ is recognized in \cite{Arbuzova:2012su} 
as fine tuning, 
the initial amplitude is found to be of
order $\delta \xi=\kappa \,(2n+1)^{3/2}$. 
Putting all things together we obtain the initial energy density of
the scalaron condensate:
\be 
\label{density}
\rho_\phi(t=0)= \la\phi^2\ra\, \omega^2(t=0)
=9\lambda^2 n^2 (2n+1)^2 \,M_P^2 H_0^2\,
\l\frac{2\Omega_{\Lambda}}{\lambda}\r^{4n+2} \l\frac{\rho_c}{\rho_{m0}}\r^{4n+1}.
\ee
Accounting for the Friedman equation \eqref{Friedman} we estimate for the
set of parameters considered in \cite{Arbuzova:2012su} ($n=2$, $\lambda=1$,
$\kappa=0.04$) the initial scalaron energy density 
\be 
\rho_\phi(t=0)= 1.5 \times 10^{-4}\,\rho_c\;,
\ee
that actually exceeds the radiation (CMB) energy density at
present. However one 
cannot expect such a large contribution of scalarons because less than one particle inside the horizon may
be created in the present (or recent) Universe, see \cite{Gorbunov:2014fwa}.
Scalarons created in the very
early Universe were very heavy (with mass $m$) 
and hence decayed to the SM particles. 
So the initial conditions in \cite{Arbuzova:2012su} seem to be
irrelevant.

\section{Relevant initial conditions for scalaron field}
\label{Sec:IC}

When density changes in a contracting object the form of scalaron
potential changes: its minimum goes closer to $\phi=0$ and its mass
rises. The initial condition $y_0'=0$ in \cite{Arbuzova:2013ina}
implies that scalaron at $t=0$ is put into the moving minimum with zero
'velocity'. But actually we should expect that at $t<0$ (i.e. before
the contraction starts) there were no excitations and scalaron was in
the vacuum state. Also we should propose that in the real situation
the contraction starts in a smooth way providing the adiabatic evolution
near $t=0$ \cite{Gorbunov:2014fwa} 
and oscillations should be excited with the
minimal possible amplitude. In what follows it is convenient to
introduce dimensionless
time $\tau=t/(\kappa t_J)$, where $t_J$ is the Jeans time, and new
variables 
\be
\label{Var}
\bar{\xi}=\xi-\xi_{min}\,,
\ee
where
\be 
\label{xi-min}
\xi_{min}=(1+\kappa \tau)^{-(2n+1)} \,.
\ee 
So the {\it adiabatic} solution of the
scalaron equation of motion derived in \cite{Arbuzova:2012su}, 
\be 
\label{eom}
\bar{\xi}''+\Omega^2 \bar{\xi}= -\xi_{min}''\,,
\ee
with such (zero) initial conditions that at $t=0$ one has $\bar{\xi}=0$,
$\bar{\xi}'=0$ looks to be the closest to the realistic 
physical situation. Here
we use notations of paper \cite{Arbuzova:2012su}, 
\be
\label{eq:Omega}
\Omega\equiv\frac{(1+\kappa \tau)^{n+1}}{\sqrt{2n+1}},
\ee
and primes correspond to derivatives with respect to $\tau$. 
\footnote{We treat maximal scalaron
   mass $m$ as being very large compared to the scale of cosmological
   constant, so the parameter $g$ used in \cite{Arbuzova:2012su} may be
   set to zero in this case.}

Note that the source in the r.h.s. of \eqref{eom} even with zero 
initial conditions 
leads to oscillations of $\bar{\xi}$. The adiabatic solution of
\eqref{eom} may be obtained by the standard technique in the form 
\be
\label{sol} 
\bar{\xi}=-\xi_1 \,\int \xi_2 \,\xi_{min}''\, d\tau+ \xi_2 
\,\int \xi_1\, \xi_{min}'' \,d\tau\,,
\ee
where 
\be 
\xi_1=\frac{1}{\sqrt{\Omega}}\sin{\int \Omega d\tau}\,,~~
\xi_2=\frac{1}{\sqrt{\Omega}}\cos{\int \Omega d\tau}\,.
\ee
The solution \eqref{sol} may be rewritten in the form 
(s.f. eq.\,(12) of Ref.\,\cite{Arbuzova:2012su}) 
\be 
\bar{\xi}= \alpha(\tau)\sin{\l\int \Omega d\tau+\delta(\tau)\r}.
\ee
After some calculations one finds the amplitude $\alpha(\tau)$ 
of generated oscillations in the limit of small $\kappa\ll 1$, 
\be 
\label{initial}
\alpha(\tau) \simeq C_{n} \kappa^{2} 
(1+\kappa \tau)^{-\frac{n+1}{2}},~~C_n=2(n+1)\,(2n+1)^{2}\,.
\ee

This result corresponds to the initial amplitude of oscillations equal
to $\alpha_0=C_n \kappa^2$, not of order $\kappa$ as it is proposed in
\cite{Arbuzova:2012su}.  We obtain parametrically smaller amplitude
for matter densities much larger than the critical one
$\rho_{m0}\gg\rho_c$ which are relevant for the astrophysical
processes (star and galaxy formation). In Appendix we find the
oscillations amplitude for more realistic Tolman model of spherical
contraction in the expanding Universe. If the initial conditions are
set in the early Universe, well before the moment when the contraction
starts, one obtains much stronger (exponential) suppression of the
scalaron oscillations for small enough $\kappa$. The evolution of the
scalaron turns to be adiabatic in this limit.

However, for matter
densities close to the critical one (cosmological processes),
$\rho_{m0}\sim\rho_c$, the amplitude of oscillations may still be large, see
eq.\,\eqref{field}, providing a significant contribution to
the energy density of the Universe\,\eqref{density} 
for some choice of parameters:
e.g., $n=2$ and $\lambda\approx 1$. In this region of model parameter
space the detailed analysis of theoretical consistency and
phenomenological and cosmological viability of $F(R)$-model is needed.

As noted in \cite{Arbuzova:2012su, Arbuzova:2013ina} scalaron
oscillations may produce massless particles. But
when $\xi>0$  (regular region, see eqs.\,\eqref{y_eom}, 
\eqref{xi_eom}) the process cannot be efficient because of very
small effective scalaron mass. We show that using 
the approximation of decaying scalaron condensate 
which works perfectly well in the regular region. The scalaron 
density for the amplitude \eqref{initial} is of order 
\be 
\rho \sim M_P^2 H_0^2 \l\frac{\rho_{m0}}{\rho_c}\r^{-6n-2}.
\ee
We use the scalaron decay width to (massless) gauge bosons from
\cite{Gorbunov:2012ns}, $\Gamma\sim\omega^3/M_P^2$, and estimate the
energy density of particles created by scalaron oscillations during the
Universe lifetime as 
\be 
\rho_p \sim \frac{\rho\,\Gamma}{H_0}=H_0^4\,\l\frac{\rho_{m0}}{\rho_c}\r^{-3n+1},
\ee
which in any case means less than one particle in a horizon-size region. 
The reason here is clear, since $\omega\sim H_0$, the scalarons are
stable at the cosmological time-scale, $\Gamma\sim H_0(H_0/M_P)^2\ll
H_0$. 
A similar result has been obtained in \cite{Arbuzova:2013ina}.

  Only if $\xi$ becomes negative (in
a spike region as it is called in \cite{Arbuzova:2013ina}) its
effective mass may increase because the scalaron potential at
negative $\phi$ may be very steep. 
One observes from \eqref{old-initial}, \eqref{Var}, \eqref{xi-min}, 
\eqref{initial}, that the spike region can be reached on the
Jeans time scale only for matter densities $\rho_m$ 
close to $\rho_c$, and hence is relevant only for 
the recent cosmic structure formation.

In the next Section\,\ref{Sec:Spike} 
we estimate the flux of particles produced
by scalaron oscillations in the spike region ($\xi<0$).

\section{Particle production in the spike region}
\label{Sec:Spike}

Hereafter we consider the case of matter densities $\rho_m$  close to $\rho_c$
which may help $\xi$ to reach negative values. In this region of
parameters the time of contraction is very large, of order of the
Universe age and the contraction has been started not long ago. So we
are interested in the evolution only until $t_J$, and for
that time one has $\Omega \kappa t\lesssim 1$, see
   Eq.\,\eqref{eq:Omega},  and 
$\xi$ reaches negative values only once or twice each time
producing a spike in the solution for curvature \cite{Arbuzova:2012su,
   Arbuzova:2013ina}.  Below we estimate the particle production
during one spike.

Homogeneous oscillations of classical scalaron field may produce
non-conformal particles and production of scalars
minimally coupled to gravity is the most efficient
\cite{Gorbunov:2010bn}. Scalaron is coupled to scalar field
$h$ through the kinetic mixing in lagrangian \cite{Gorbunov:2010bn} 
\be 
L_{int}=\frac{h}{\sqrt{6}M_P}\d_{\mu}\phi\, \d^{\mu} h\,.
\ee
This coupling modifies the equation of motion for $h$:
\be 
\label{h-production}
\d_{\mu}\d^{\mu} h + \l \frac{\ddot{\phi}}{\sqrt{6}M_P}+m_{h}^2\r h=0\,,
\ee
where $m_h$ is the scalar mass. Scalaron field here plays a role of
the external force producing particles $h$. The number of produced
particles may be estimated by making use of the Bogoloubov
transformations. From eq.\,\eqref{h-production} one obtains for the
Fourier mode $h_k$ of 3-momentum $k$, 
\be 
\label{mode}
\ddot{h}_k+\l k^2+ m_{eff}^2(t) +m_{h}^2\r h_k=0\,,
\ee
where $m_{eff}^2(t)=\ddot{\phi}/(\sqrt{6} M_P)$.

Note that the spikes correspond to the negative values of $\phi$ and
in that region the scalaron mass is maximal, 
so one has $\ddot{\phi}=-m^2\phi$ during the spike. The maximum value of
$|\phi_{m}|$ may be extracted from the maximum $|\xi_{m}|$ estimated
in \cite{Arbuzova:2012su}: 
\be \frac{|\phi_{m}|}{\sqrt{6}M_P}=N
\frac{H_0}{m}\l\frac{\rho_{m0}}{\rho_c}\r^{-n}z_1^{-n},
~~N=\sqrt{\frac{\lambda}{2}}\l
\frac{2 \Omega_{\Lambda}}{\lambda} \r^{n+1/2}. 
\ee 
Here $z_1\equiv 1+t_1/t_J$
where $t_1$ corresponds to the moment when $\xi$ crosses zero
for the first time: $\alpha(\tau_1)=\xi_{min}(\tau_1)$. Note
that for the case we consider the variable $z$ is in the interval 
$1<z_1<2$,  so we put it to be
$z_1=1$ hereafter having in mind an upper bound. 
Then the maximum possible value of $m_{eff}^2(t)$ is
\be 
M^2=N m H_0 \l\frac{\rho_{m0}}{\rho_c}\r^{-n}.
\ee

  If we approximate the spike by the gaussian form with height of
   $M$ and width of $1/m$ we obtain formally that the spectrum has a
   cut off at high momenta of order $m$. The calculation shows that
   high energy particles with number density of order $n_p\sim M^4/m$
   are produced by spike, in agreement with the statement of
   Ref.\,\cite{Arbuzova:2012su}.

Since production of such energetic particles in a slow and smooth process
of contraction looks very surprising from any point of view it is
needed to check the validity of all approximations in use. 
The approximation that is certainly 
doubted is the homogeneous energy-momentum tensor for the background matter. If one considers a set of discrete particles instead of continuous medium one obtain that they
produce the scalaron field like point sources. Every particle
produces the field $\xi\propto e^{-\omega r}/r$ like the usual massive
scalar with mass $\omega$. If the average distance $d$ between particles
is large, $d\gg 1/\omega$, then the scalaron field is strongly inhomogeneous and
approximation used for the energy-momentum tensor is not valid.

In this case we expect that the energetic particle
production may still take place only in the small vicinities of the point
sources, i.e. in very small spherical regions of radius 
$1/m$. Therefore, the total number of produced particles is 
suppressed by a very small number $n/m^3$ where $n$ is the matter number
density. For the energy density we get the estimate
\be 
\label{CR-density}
\rho \sim
\frac{M^4 \,n}{m^3}=N^2 \rho_{m0}
\,\frac{H_0^2}{m_0\,m}\l\frac{\rho_{m0}}{\rho_c}\r^{-2n}, 
\ee 
where $m_0$ is the mass of particle populating the contracting object
(region). For the interesting astrophysical objects and any regions
dominated by baryons one puts $m_0\sim1$\,GeV. Then for the 
close to critical density and the
scalaron mass $m\sim 10^{13}$ GeV we obtain $\rho\sim 10^{-92}~ {\rm
   GeV/(cm^2\,s)}$ which is too small number to talk about.

Note in passing, though we discussed the particle production from one
spike, the consideration of a sequence of several spikes does not
significantly change the result. Another comment concerns the
contraction of dark matter dominated region. To the case of
particle-like dark matter (e.g. weakly interaction massive particles)
the estimate \eqref{CR-density} is applicable with $m_0$ referring to
the dark matter particle mass. Say, for the galactic dark matter
particle heavier than 1 eV the size of its wave packet (de Broglie
wavelength corresponding to the average velocity $10^{-3}$) is smaller
than the distance between particles, hence the homogeneous description
fails.  This problem is expected to be generic for $F(R)$ gravity
models constructed to describe the dark energy. Typically, the
effective scalaron mass can vary with the background matter density by
many orders of magnitude. This happens because $F(R)$ functions
contain two different mass scales: the present day Hubble parameter
and the maximal scalaron mass which must be introduced to avoid
singularities \cite{Appleby:2009uf}. The latter is bounded from below
as $m\gtrsim 10^5$ GeV \cite{Arbuzova:2011fu}. Therefore, the scalaron
degree of freedom can possess the similar dynamics including the
non-linear oscillations (spike region). In this region the validity of
the homogeneous description is questioned. Note that the same problem
is also inherent in chameleon models
\cite{Khoury:2013yya,Burrage:2017qrf}.

The case of homogeneous oscillating scalar field (e.g. axion-like)
playing the role of the dark matter is actually inconsistent with the
stability of the studied in \cite{Arbuzova:2012su, Arbuzova:2013ina}
$F(R)$-gravity \eqref{FR}. This model is known to be well defined only
for $R\gtrsim R_0$ \cite{Appleby:2009uf}. For lower values of $R$ the
scalaron degree of freedom behaves as a tachyon. If the scalar field
condensate dominates than the trace of the energy momentum tensor
oscillates reaching negative values bringing the curvature to the
region of instability. The same problem arising at the stage of
inflaton oscillations was considered in detail in
\cite{Appleby:2009uf} (see Sec. 4.1 of that paper). This problem can
be avoided with an appropriate choice of the function $F(R)$. However,
we expect the scalaron dynamics to be strongly affected by the
oscillating condensate. The corresponding effects need a special study
which is beyond the scope of this paper. To conclude, the homogeneous
approximation for the matter, as well as the results of
\cite{Arbuzova:2012su, Arbuzova:2013ina} for the efficient cosmic ray
production in the spike region, are invalid for a generic dark matter
model of this type.

To summarize, we have shown that contracting objects made of baryons
or dark matter particles in $F(R)$-gravity do not contribute to the cosmic ray
spectrum via the decay of the scalaron oscillations.

\vspace{0.3cm}
We thank  E.Arbuzova, F.Bezrukov, A.Dolgov, D.Levkov, A.Panin,
V.Rubakov and S.Sibiryakov for correspondence and discussions. 
The work has been supported by Russian Science Foundation grant 14-12-01430.

\appendix

\section{The realistic model for the Jeans contraction during the structure formation}

A structure formation process starts at the matter dominated stage. In
the linear regime small perturbations grow as $\delta\rho/\rho\propto
a\propto t^{2/3}$. When $\delta \rho/\rho$ reaches unity the
non-linear contraction starts. For the spherically symmetric
perturbations, the smooth transition between this two regimes may be
approximately described by the Tolman solution \cite{Mukhanov:2005sc}:
\be 
\label{tolman}
t=\sqrt{\frac{3}{8}} t_{J} (\zeta+\sin{\zeta}),~~\rho_m=\frac{8 \rho_{m0}}{(1+\cos{\zeta})^3}~.
\ee
Here $t_J=1/\sqrt{4\pi G\rho_{m0}}$. The moment $t=0$ correspond to
the time when the density starts to grow. This solution has two
singularities: in the past and in the future. The singularity in the
past corresponds to the Big Bang while the singularity in the future
at $\zeta=\pi/2$ is never reached for real objects. Clearly, the
solution is valid only for $t\lesssim t_J$.

The motivated initial conditions on the scalaron field must follow
from the fact that all scalarons (if ever produced) had been decayed
well before the matter dominated stage started. Therefore, the scalaron had
been placed to the minimum and moved together with it
adiabatically before the Jeans contraction started. Here we show that
the resulting amplitude of scalaron oscillations is strongly
suppressed compared to the case when the vacuum initial conditions are
set at the moment when the contraction starts.

\begin{figure}[!htb]
  \centerline{
    \includegraphics[width=0.3\textwidth]{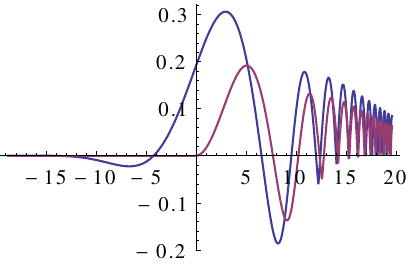}
    \hskip 0.03\textwidth
    \includegraphics[width=0.3\textwidth]{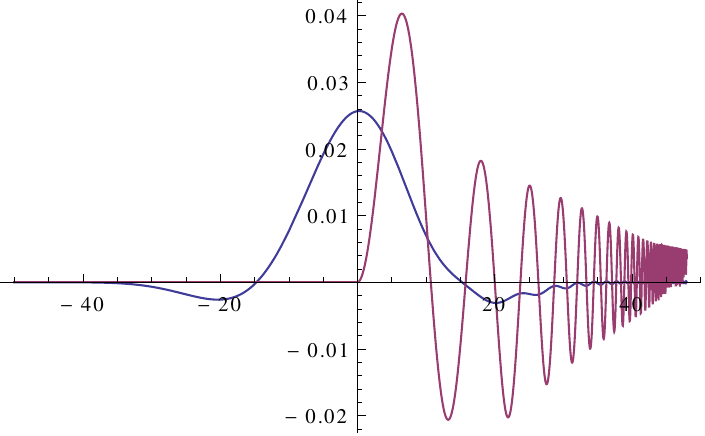}
    \hskip 0.03\textwidth
\includegraphics[width=0.3\textwidth]{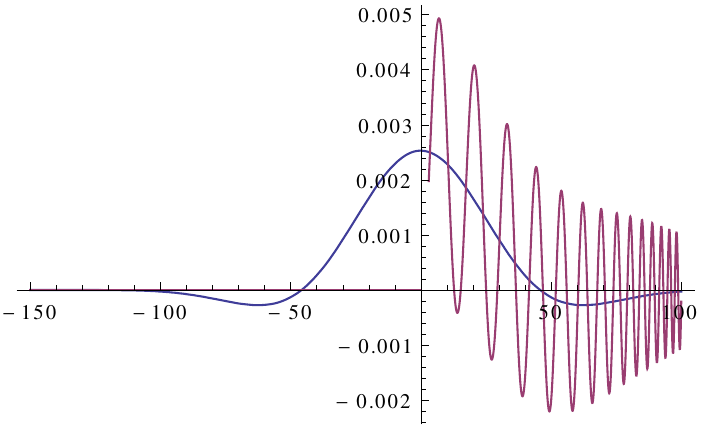}
}
\caption{\label{Fig1} 
  These plots show the evolution of $\xi-\xi_{min}$ for
  $\kappa=0.08$ (left), $\kappa=0.03$ (middle) 
and $\kappa=0.01$ (right). The blue line corresponds to initial
  conditions 
$\xi(\tau_0)=\xi_{min}(\tau_0)$, $\xi'(\tau_0)=\xi_{min}'(\tau_0)$
  imposed at $\tau_0=-1.5/\kappa$, 
  the violet line is for $\tau_0=0$. Taking the earlier
  initial time for the first choice of conditions provides
  with the same result.}
 \end{figure}

We numerically solve both equations \eqref{xi_eom} and \eqref{y_eom}
for the Tolman model of density evolution \eqref{tolman}. As it is done
in \cite{Arbuzova:2012su}, we introduce the dimensionless parameter
$\kappa =1/(m\sqrt{g} t_J)$ and present plots for several choices of
$\kappa$, see Fig.\,\ref{Fig1}.
The
larger $\kappa$ correspond to the smaller minimal densities $\rho_{m0}$.
 
One observes in Fig.\,\ref{Fig1}
that for relatively large realistic values of $\kappa$ we
obtain larger amplitude of oscillations than for setting initial
conditions at $\tau_0=0$. But for small $\kappa$ the late
time oscillation amplitude becomes negligible compared to the case
$\tau_0=0$. For the case $\tau_0=0$ we obtain that the final amplitude
of induced oscillations behaves as $\kappa^2$ similarly to the results
of the Sec.\,\ref{Sec:IC}.
From Fig.\,\ref{Fig1} (middle and right panels) it is seen
that for more realistic choice of $\tau_0\sim-1/\kappa$ the amplitude
drops with $\kappa$ much faster.

The exponential damping of the
discussed effect for the small scale structures with densities much
larger than the critical one can be understood analytically. 
The solution \eqref{sol} of the linearised equation \eqref{eom} can be
rewritten in the form
\be 
\bar{\xi}=A(\tau)\sin{\l\int \Omega d\tau+\delta(\tau)\r},
\ee
where $\Omega$ is approximated as $\Omega=(1+\kappa^2
\tau^2/6)^{n+1}/\sqrt{2n+1}$ for $\tau<1/\kappa$.

The amplitude $A(\tau)$ can be calculated as
\be 
A(\tau)=|\int^{\tau}_{-\infty}d\tau_1\,\xi_{min}''(\tau_1)\frac{e^{i \int_0^{\tau_1} \Omega d\tau_2}}{\sqrt{\Omega}}|
\ee
The oscillations are generated near $\tau=0$. For $\tau>1/\kappa$ the
impact of the source $\xi_{min}''$ becomes negligible so the amplitude
is actually the same as in the formal limit $\tau\rightarrow
\infty$. The latter is given by the integral
\be 
A=|\int^{\infty}_{-\infty}d\tau_1\, \xi_{min}''(\tau_1)\frac{e^{i \int_0^{\tau_1} \Omega d\tau_2}}{\sqrt{\Omega}}|=|\int^{\infty}_{-\infty}d\tau_1\,\frac{e^{i \int_0^{\tau_1} \Omega d\tau_2} P(\kappa \tau)}{(1+\kappa^2 \tau^2/6)^{\alpha}}|\,,
\ee
where $P(\kappa \tau)$ is some polynomial function, $\alpha=(5
n+7)/2$. Calculating this integral with residues one finds the main
exponential dependence:
\be 
A\propto e^{-\frac{\beta(n)}{\kappa}}, \beta(n)=\frac{\sqrt{6 \pi}\,\Gamma(n+2)}{2\sqrt{2n+1}\,\,\Gamma(n+5/2)}\,.
\ee
Thus, for small $\kappa$ oscillations are strongly suppressed, in
agreement with the numerical calculations presented in Fig.\,\ref{Fig1}.



\begin{thebibliography}{99}

\bibitem{Starobinsky:2007hu}
   A.~A.~Starobinsky,
   JETP Lett.\  {\bf 86}, 157 (2007)
   [arXiv:0706.2041 [astro-ph]].

\bibitem{Nojiri:2006ri}
   S.~Nojiri and S.~D.~Odintsov,
   eConf C {\bf 0602061} (2006) 06
    [Int.\ J.\ Geom.\ Meth.\ Mod.\ Phys.\  {\bf 4} (2007) 115]
   [hep-th/0601213].

\bibitem{Sotiriou:2008rp}
   T.~P.~Sotiriou and V.~Faraoni,
   Rev.\ Mod.\ Phys.\  {\bf 82}, 451 (2010)
   [arXiv:0805.1726 [gr-qc]].

\bibitem{Nojiri:2010wj}
   S.~Nojiri and S.~D.~Odintsov,
   Phys.\ Rept.\  {\bf 505} (2011) 59
   [arXiv:1011.0544 [gr-qc]].

\bibitem{Arbuzova:2012su}
   E.~V.~Arbuzova, A.~D.~Dolgov and L.~Reverberi,
   Eur.\ Phys.\ J.\ C {\bf 72}, 2247 (2012)
   [arXiv:1211.5011 [gr-qc]].

\bibitem{Arbuzova:2013ina}
   E.~V.~Arbuzova, A.~D.~Dolgov and L.~Reverberi,
   Phys.\ Rev.\ D {\bf 88}, no. 2, 024035 (2013)
   [arXiv:1305.5668 [gr-qc]].

\bibitem{Gorbunov:2014fwa} 
  D.~Gorbunov and A.~Tokareva,
  J.\ Exp.\ Theor.\ Phys.\  {\bf 120}, no. 3, 528 (2015)
  doi:10.1134/S1063776115030085
  [arXiv:1412.3413 [astro-ph.CO]].

\bibitem{Gorbunov:2011zzc}
   D.~S.~Gorbunov and V.~A.~Rubakov,
   Hackensack, USA: World Scientific (2011) 489 p

\bibitem{Arbuzova:2010iu}
   E.~V.~Arbuzova and A.~D.~Dolgov,
   Phys.\ Lett.\ B {\bf 700}, 289 (2011)
   [arXiv:1012.1963 [astro-ph.CO]].

\bibitem{Gorbunov:2012ns}
   D.~Gorbunov and A.~Tokareva,
   JCAP {\bf 1312}, 021 (2013)
   [arXiv:1212.4466 [astro-ph.CO]].

 \bibitem{Gorbunov:2010bn}
   D.~S.~Gorbunov and A.~G.~Panin,
   Phys.\ Lett.\ B {\bf 700}, 157 (2011)
   [arXiv:1009.2448 [hep-ph]].



\bibitem{Appleby:2009uf}
   S.~A.~Appleby, R.~A.~Battye and A.~A.~Starobinsky,
   JCAP {\bf 1006}, 005 (2010)
   [arXiv:0909.1737 [astro-ph.CO]].
   
\bibitem{Arbuzova:2011fu}
   E.~V.~Arbuzova, A.~D.~Dolgov and L.~Reverberi,
   JCAP {\bf 1202}, 049 (2012)
   [arXiv:1112.4995 [gr-qc]].

\bibitem{Khoury:2013yya} 
  J.~Khoury,
  Class.\ Quant.\ Grav.\  {\bf 30}, 214004 (2013)
  [arXiv:1306.4326 [astro-ph.CO]].
  
\bibitem{Burrage:2017qrf} 
  C.~Burrage and J.~Sakstein,
  arXiv:1709.09071 [astro-ph.CO].
  
\bibitem{Mukhanov:2005sc} 
  V.~Mukhanov,
  ``Physical Foundations of Cosmology'',
   Cambridge University Press (2005)
\end{thebibliography}
\end{document}